\documentclass[preprint,sort&compress,3p]{elsarticle}
\usepackage{amsmath}
\usepackage{amsfonts}
\usepackage{amssymb}

\usepackage{subfig}
 
\usepackage{cleveref}

\usepackage[usenames,dvipsnames]{color,colortbl}

\usepackage[export]{adjustbox}[2011/08/13]

\usepackage[figurename=Fig.,labelfont=bf,tablename=Tab.]{caption}
\DeclareCaptionLabelSeparator{bar}{\hspace{0.3mm}$\boldsymbol{|}$\hspace{0.3mm}}
\usepackage{listings}
\definecolor{codegreen}{rgb}{0,0.6,0}
\definecolor{codegray}{rgb}{0.5,0.5,0.5}
\definecolor{codepurple}{rgb}{0.58,0,0.82}
\definecolor{backcolour}{rgb}{0.95,0.95,0.92}
\definecolor{codedarkgreen}{rgb}{0.263,0.537,0.345}
\definecolor{codedarkblue}{rgb}{0.235,0.49,0.568}
\definecolor{mathcomments}{rgb}{0.271,0.580,0.682}
\definecolor{Gray}{gray}{0.9}
\lstdefinestyle{mystyle}{
    backgroundcolor=\color{backcolour},   
    commentstyle=\color{codegreen},
    keywordstyle=\color{magenta},
    numberstyle=\tiny\color{codegray},
    stringstyle=\color{codepurple},
    basicstyle=\ttfamily\footnotesize,
    breakatwhitespace=false,         
    breaklines=true,                 
    captionpos=b,                    
    keepspaces=true,                 
    numbers=left,                    
    numbersep=5pt,                  
    showspaces=false,                
    showstringspaces=false,
    showtabs=false,                  
    tabsize=2,
    escapechar={|}
}
\begin{document}
\begin{frontmatter}
\journal{ }
\title{Phase-field modelling of failure in ceramics with multiscale porosity}
\author[1]{R. Cavuoto} 
\author[2]{P. Lenarda}
\author[3]{A. Tampieri}
\author[4]{D. Bigoni\corref{cor}}
\ead{bigoni@ing.unitn.it}
\author[2]{M. Paggi\corref{cor}}
\ead{marco.paggi@imtlucca.it}

\address[1]{School of Engineering, University of Naples \lq \lq Federico II", Naples, Italy}
\address[2]{IMT School for Advanced Studies Lucca, Piazza San Francesco 19, 55100 Lucca, Italy}
\address[3]{Institute of Science and Technology for Ceramics, National Research Council, Via Granarolo 64, 48018 Faenza, Italy}
\address[4]{Instabilities Lab, University of Trento, Via Mesiano 77, Trento, 38123 Italy}
\cortext[cor]{Corresponding authors:}

\begin{abstract}

Many stiff biological materials exhibiting outstanding compressive strength/weight ratio are characterized by high porosity, spanning different size-scales, typical examples being bone and wood. A successful bio-mimicking of these materials is provided by a recently-obtained apatite, directly produced through a biomorphic transformation of natural wood and thus inheriting its highly hierarchical structure. This unique apatite (but also wood and bone) is characterized by two major distinct populations of differently-sized cylindrical voids, a porosity shown in the present paper to influence failure, both in terms of damage growth and fracture nucleation and propagation. This statement follows from failure analysis, developed through in-silico generation of artificial samples (reproducing the two-scale porosity of the material) and  subsequent finite element modelling of damage, implemented with phase-field treatment for fracture growth. 
It is found that  small voids promote damage nucleation and enhance bridging of macro-pores by micro-crack formation, while macro-pores influence the overall material response and drive the propagation of large fractures. 
Our results explain the important role of multiscale porosity characterizing stiff biological materials and lead to a new design paradigm, by introducing an in-silico tool to implement bio-mimicking in new artificial materials with brittle behaviour, such as carbide or ceramic foams.
\end{abstract}

\end{frontmatter}

\section{Introduction}
\label{Introduction}
\begin{figure}[ht]
 \centering
 \includegraphics[width=\linewidth,angle=0]{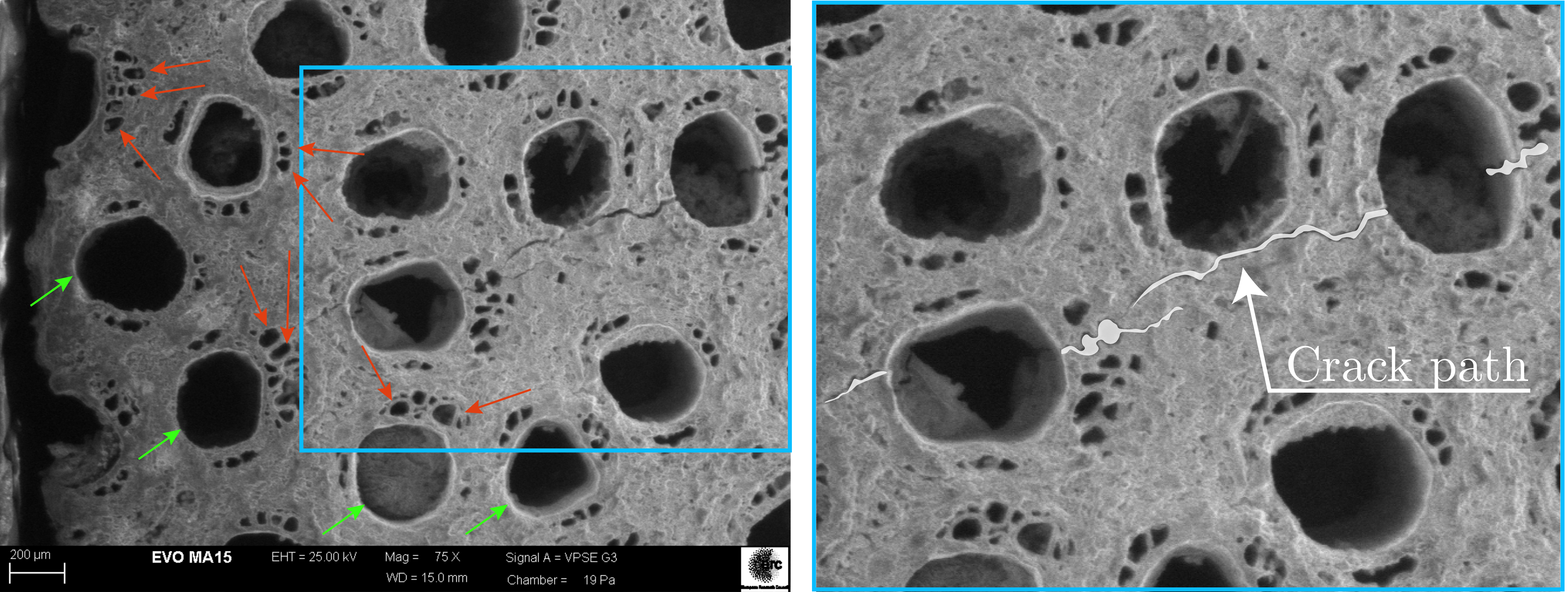}
 \caption{\textbf{Crack propagation in Biomorphic Apatite, characterized by a multiscale porosity.} Left: SEM image of a BA sample 
 characterized by the presence of macro (green arrows) and meso (red arrows) cylindrical pores with a nearly circular cross-section.
 The sample was subjected to uniaxial compression test.
 Right: a magnified detail of the sample on the left shows a 
  crack (highlighted white) which nucleated, almost parallel to the direction of loading, from the external surface of a macro pore, where tensile stress locally develops, although the mean stress in the sample is compressive.
 }
 \label{BA}
\end{figure}

Materials characterised by a high-density distribution of voids
are diffused in nature (for instance bone, wood, rock --sandstone, tuff, pumice--, coal) and as man-made materials 
(for instance concrete, cellular ceramic, and carbide foam). 
Wood and bone evidence populations of cylindrical pores with nearly circular cross-sections, ranging in diameter through multiple size scales, a feature  
affecting many of their physical \cite{sprio,wang2022,wolfel2022} and mechanical \cite{somba2021,heath2016,ozen2022} properties.
Implementing size-scale variation in the porosity of artificial materials is a bio-mimicking challenge, particularly for bone substitutes and scaffolds 
\cite{gupta2021,baux2019, senhora2022,gomez2021}.
A successful answer to this challenge was provided  \cite{tampieri} through the synthesis of a new material, 
a 3-D porous apatite (called \lq Biomorphic Apatite', BA in the following) 
obtained via a bio-morphic transformation of natural
wood into ceramics, Fig. \ref{BA}. 

BA replicates the multiscale hierarchical structure of wood, with a double-sized porosity (Fig. \ref{BA}), sharing similarities with bone and thus becoming ideal for bio-mechanical applications, where outstanding strength and fracture energy absorption are crucially important \cite{sprio,bigoni}. Here, a correct failure analysis may lead to the design of surgical interventions and prediction of rehabilitation times so that the development of a numerical tool for this analysis is the object of the present paper. A class of materials (including those addressed here, plus glass, ceramic, rock and concrete) are brittle, so that fracture nucleates primarily under tensile stresses. Under overall compression, crack generation is complicated by the presence of voids, which concentrate tensile stress near their boundary and 
 stimulate fracture growth parallel to the loading direction \cite{Roberts2000,Pabst2005,sammis}, see Fig. \ref{BA}. 
 Fracture patterns are typically tortuous and difficult to follow or model analytically \cite{kachanov}, so that numerical techniques play a decisive role. These can be classified as:  
(i.) augmented versions of the finite-element method \cite{gustaf2020}  (such as cohesive-zone methods, CZM, or extended finite-element methods, XFEM, the former requiring an initial knowledge of the crack pattern and the latter an {\it ad hoc} constitutive relation at the crack tip);
(ii.) diffusive damage models
\cite{kart2022,bazant1990,bazant1994,bui2022}
(relying on a regularized crack topology); (iii.) the phase-field approach, overcoming  the difficulties encountered in the other techniques by allowing damage to grow within a thin portion of the material. 
The latter technique has been developed as an energy minimization, via $\Gamma$-convergence regularization of free discontinuities to model cracks \cite{Francfort1998,Bourdin2008,dalMaso2002,Ambrosio1990,Maso1993,Braides1998, Miehe2010,geers1998}
and was recently proven to be particularly suitable for the description of complex fracture patterns in PMMA samples with notches and holes \cite{Bigoni2022} and therefore is adopted here. 

The porosity typical of biomorphic apatite is primarily double sized, so that  
macro-scale pores can be differentiated from meso-scale pores, Fig. \ref{BA} (micropores, also present, are neglected as they were found not to influence fracture) and is shown to correspond to two peaks of a probability density function. 
This function allows the set-up of a numerical tool for the generation of in-silico porous samples, to be used for subsequent fracture simulation using the phase-field technique. 

The simulations show that the porosity is always connected to a reduction of overall stiffness and strength of the material. However, the meso-scale porosity produces: (i.) a smearing of damage through zones that would remain intact in the absence of this scale of pores; (ii.) a shielding of regions of the material from cracks; and (iii.) a promotion of fracture growth in other zones. 

Our results show that, not only the value of \lq total porosity' is relevant to the mechanical modelling of porous-brittle solids, rather the amount of fractions of porosity with different characteristic size plays a strong role, a result with implications in the simulation of fracture in biological porous materials, such as bone or wood, and in porous ceramics for bone repair or for filters used in industrial applications.

\section{Results}
\label{ch:phase}
\subsection{Theory}

\noindent In a purely isothermal formulation of a brittle-damaging material, where the damage only occurs for tensile strains, the Helmholtz free energy per unit volume can be represented as
\begin{equation}
\label{pera}
    \psi(\mathbf{\varepsilon}, s) 
    = \left[(1-s)^2+k\right] \, \psi_+(\mathbf{\varepsilon}) +
    \psi_-(\mathbf{\varepsilon})
    , 
\end{equation}
a function of a damage parameter $s\in[1-\sqrt{1-k},1]$,
where $k>0$ represents a residual stiffness. 
Note that the lower extremum of the range of variation for $s$ corresponds to an intact material, while $s=1$ to a fully damaged one. 
Equation \eqref{pera} embodies an  additive split of the elastic energy $\psi(\mathbf{\varepsilon})$ into tensile, 
\lq$+$', and compressive, \lq$-$', strains
\begin{equation}
\boldsymbol \varepsilon_\pm = \sum_{i=1}^3 \langle \varepsilon_i \rangle_\pm \boldsymbol e_i \otimes \boldsymbol e_i,   \end{equation}
where $\boldsymbol e_i$ and $\varepsilon_i$ ($i=1,2,3$) are the unit eigenvectors and corresponding eigenvalues of the strain, respectively, and
$\langle \cdot\rangle$ denotes 
the Macaulay bracket operator, defined for every scalar $x$ as $\langle x  \rangle_{\pm} = (x \pm | x |)/2$. 

The Helmholtz free energy, Eq.\eqref{pera}, defines a diffuse and isotropic damage model, where damage only occurs in tension, so that the stress tensor
\begin{equation}
    \mathbf{\sigma}(\mathbf{\varepsilon}, s) = \frac{\partial \psi(\mathbf{\varepsilon},s)}{\partial \mathbf{\varepsilon}} , 
\end{equation}
becomes the sum of tensile and compressive components,
$
    \mathbf{\sigma} = \mathbf{\sigma}_+ 
    +\mathbf{\sigma}_-,
$
both related to the strains through the isotropic elasticity tensor
$\mathcal{E}$ as 
\begin{equation}
\begin{array}{lr}
         \mathbf{\sigma}_+ = \left[(1-s)^2+k\right]\, \mathcal{E}\mathbf{\varepsilon}_+,
    & \,\,\,\,\,\,
    \mathbf{\sigma}_- =  \mathcal{E}\mathbf{\varepsilon}_-.
    \end{array}
\end{equation}

\noindent Following the variational approach to brittle fracture, propagation and branching of a crack in a damaging solid can be found as the result of the minimization of the energy functional \cite{Miehe2010,Borden2012} 
\begin{equation}
\Pi(\mathbf{u}, s, \Gamma) =   \int_{\Omega \backslash \Gamma } \psi(\boldsymbol \varepsilon, s)  \ \mathrm{d}  \mathbf{x} + \int_\Gamma \mathcal{G}_c(s)\ \textup{d}\Gamma,
\label{notregular}
\end{equation}
where $\mathbf{u}$ is the displacement, $\mathcal{G}_c$ is the fracture energy and $\Gamma$ is the (unknown and evolving) fracture path inside the body $\Omega$.

\noindent 
A direct use of the functional \eqref{notregular} involves the solution of a free boundary value problem, which can be approximated by following the regularised framework introduced with the phase field approach \cite{Bourdin2008,Miehe2010}. The approximation regards the topology of the crack, which is smeared out onto the whole body, allowing to rewrite the potential energy of the system as a volume integral 
\begin{equation}\label{regular}
\Pi (\mathbf{u}, s) = \int_{\Omega} \left[\psi(\boldsymbol \varepsilon,s) \,+G_c \,
 \gamma (s, \nabla s) \right]\ \mathrm{d}  \mathbf{x} \ ,
\end{equation}
where $\gamma(s, \nabla s)$ is the crack density functional, depending on the spatial gradient $\nabla \cdot $ of the  internal state variable $s$, now called \lq phase field variable'. 

\noindent 
The surface $\Gamma$, not explicitly present in the functional \eqref{regular}, can be recovered as the result of an energy minimization procedure.
In particular, introducing the so-called \lq AT2 model' \cite{Miehe2010, Ambrosio1990}, the functional $\gamma (s, \nabla s)$,  is represented by the following convex function
\begin{equation}
 \gamma(s, \nabla s) = \frac{1}{2l} s^{2} + \dfrac{l}{2} |\nabla s |^{2} \ ,
 \label{crackdensity}
\end{equation}
where $l$ is a regularisation characteristic length, related to the smeared crack width. 
For a sufficiently small $l$, the minimization of functional \eqref{regular}, 
implemented with the nonlocal crack density function \eqref{crackdensity}, 
\lq spontaneously' leads to a solution where the damage is strongly localized near a surface $\Gamma$, representing a crack. 
Formally, it can be demonstrated 
that, for vanishing regularisation parameter ($l \to 0$), the formulation outlined in Eq.\eqref{regular} tends to Eq.\eqref{notregular} in the sense of the so-called \lq $\Gamma$-convergence' \cite{Braides1998}. 

\noindent
Finally, the weak form of the variational problem, and the consequent finite-element discretization, can be found in \cite{Bigoni2022}.

\section{Virtual testing} 
\subsection{Two scale porosity and mesh generation} \label{ch:samplegen}

\noindent A brittle porous material is now considered, characterized by a two-scale cylindrical porosity. 
A two-dimensional formulation fits the specific {\it transverse} structure that biomorphic apatite (\lq BA' henceforth) inherits from 
rattan wood, which presents elongated grains and parallel channels, with approximately circular cross-section and two predominant sizes (Fig. \ref{BA}).

\begin{figure}[ht]
 \centering
 \includegraphics[width=0.6\linewidth]{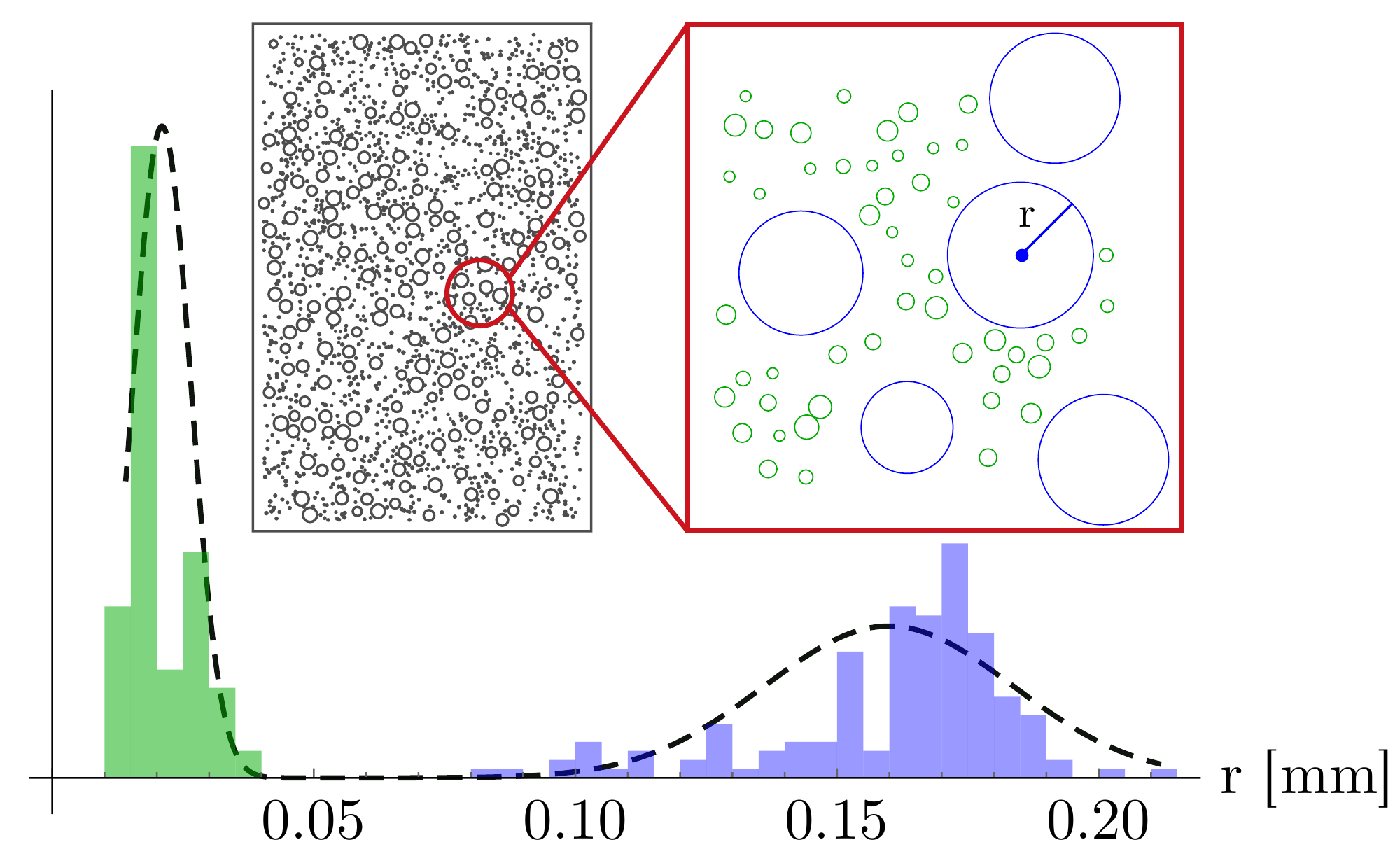}
 \caption{\textbf{Size distribution of voids in Biomorphic Apatite, characterized by multiscale porosity.} Probability density function (PDF, shown dashed) 
 obtained from SEM photomicrographs of BA samples, modelling their 
 void distribution in terms of the voids' radius, $r$. The BA specimens evidence a bimodal distribution of void radii typical of a double porosity (macro-porosity is marked in blue, while meso-porosity in green). The inset shows a sample generated in-silico with the PDF reported in Fig. (black dashed curve), which has been coupled with a random generator to impose the position of the voids' center.
 }
 \label{FIG_PDF_SAMPLE}
\end{figure}

The BA samples are characterized by three size-levels of porosity $0 \leq \Phi \leq 1$ (defined as the ratio between the volume of voids and the total volume of the element in a representative volume element), namely, macroscopic (approximately 300\,$\mu$m in diameter), mesoscropic (up to 50 $\mu$m in diameter), and microscopic (1~$\mu$m of diameter) \cite{sprio}.
Using an {\it ad hoc} developed Matlab code  
(based on image analysis functions 
\lq imread' and \lq imfindcircles\rq, available in the complementary material), the porosity  of BA, as 
characterized 
 from SEM photomicrographs, was shown to obey to the following 
 probability distribution function (PDF) 
\begin{equation}
    f = 16.82\ \mathrm{e}^{-889.1(r-0.16)^2}+72.24\ \mathrm{e}^{-16396(r-0.02)^2},
    \label{eq:PDF}
\end{equation}
which exhibits the bi-modal shape, clearly visible in Fig. \ref{FIG_PDF_SAMPLE} (black dashed curve).

The peak on the right  of the curve 
occurs at high values of void radius and is representative of  macro-porosity $\Phi_{\rm{macro}}$, while 
the peak on the left is representative of meso-porosity $\Phi_{\rm{meso}}$, so that, neglecting the micro level, the total porosity is the sum of the two following major contributions $\Phi= \Phi_{\rm{macro}} + \Phi_{\rm{meso}}$. 

The obtained PDF was used for  
in-silico generation (through random placement of non-overlapping voids) of samples of porous materials (an example is shown in the inset of Fig. \ref{FIG_PDF_SAMPLE}), needed for simulations (performed at the Laboratory for Numerical Modelling of Materials of the University of Trento, using a 
Workstation AMD Ryzen Threadripper PRO 5995WX, 512GB RAM, 2 GeForce RTX 3090 graphic cards, acquired with the ERC-AdG-2021-101052956-Beyond), via  the previously outlined phase-field approach. 

In order to investigate the influence of two size-scale porosity on crack patterns in ceramic materials subject to uniaxial compression, a 
preliminary calibration (reported in the supporting material) on single-porosity samples [generated by simply neglecting the right term in equation (\ref{eq:PDF})] has been performed for both mechanical model and  mesh size for subsequent finite element solution. Results of this analysis are shown in the supporting material.

For a material characterized by a porosity $\Phi$, independently of the number of size-scales of porosity, the 
overall stiffness of an equivalent homogeneous material can theoretically be estimated in a number of ways. Following the Coble-Kingery model \cite{COBLE1956,PABST2003}, the effective Young's modulus $\bar{E}$, defining a homogeneous elastic solid equivalent to the porous material, is 
\begin{equation}
\label{eq:coble}
\frac{\bar{E}}{E_{\textup{matrix}}}=(1-\Phi)^m\ ,
\end{equation}
where $m=2$ is to be chosen, for spherical voids and Poisson's ratio close to 0.2. The Hashin-Shtrikman upper bound \cite{HS1963} was also used, see supporting material. 

The peak stress under increasing uniaxial stress $\bar{\sigma}$ (corresponding to failure in a force-controlled testing device), for a sample made of an equivalent homogeneous material, 
is assumed to follow the simple rule  \cite{rice1998} 
\begin{equation}
\label{eq:strength}
\frac{\bar{\sigma}}{\sigma_{\textup{matrix}}}=\textup{e}^{-b\Phi},
\end{equation} 
where $\sigma_{\textup{matrix}}$ is the peak (or failure) stress in the matrix material and $b$ a coefficient to be fitted with experimental measurements.
\sloppy

Equations \eqref{eq:coble} and \eqref{eq:strength} are used as reference in the simulations, in particular, once the Young's modulus and failure stress of the matrix material are assigned for numerical calculations, the above equations are employed to obtain an expected \lq rough' behaviour 
of porous samples.

\begin{figure}[ht]
\centering
\includegraphics[width=1.0\linewidth]{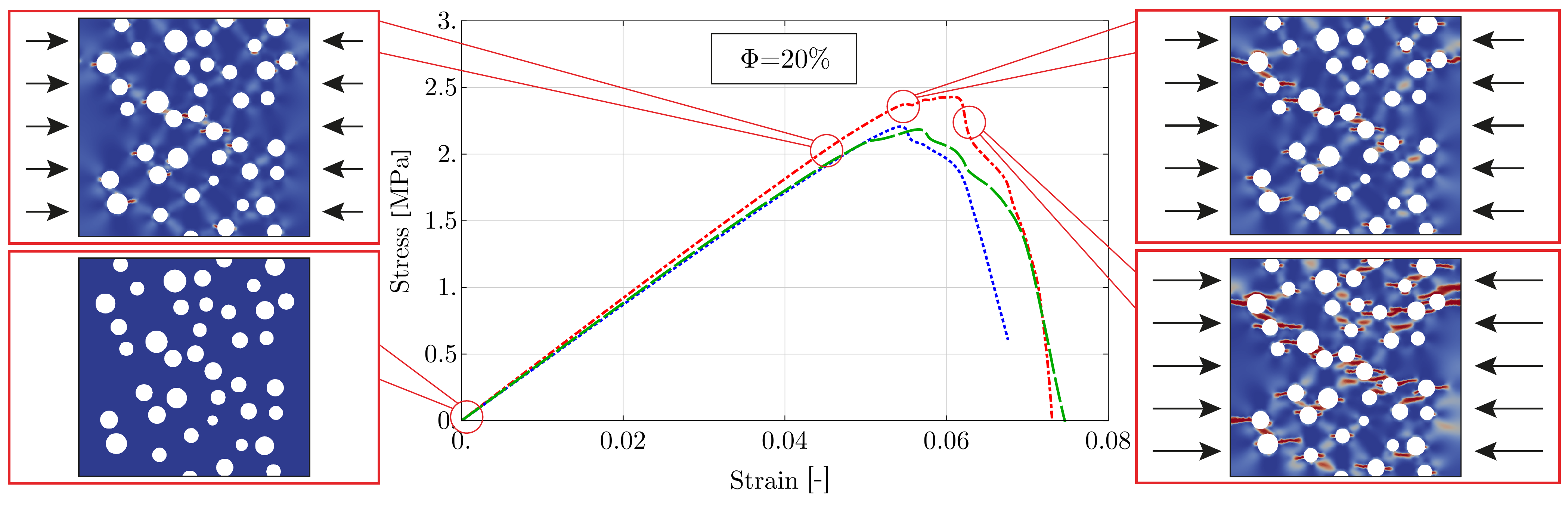}
	\caption{\textbf{Compression behaviour of 
 a simulated biomorphic apatite, 
 where only one scale of porosity is considered, using phase-field.} Simulated overall stress-strain responses for three samples with the same porosity $20\%$, but different void distributions. The progressive crack formation and growth, developing parallel to the load direction, is  reported at increasing values of overall strain $\bar{\varepsilon}$ in the inset for the sample with  $\Phi_{\textup{macro}}=20\%$ shown unloaded in Extended Data of the supporting material.
Note the strong effect of crack propagation on the overall stress-strain curves. 
}
\label{StressvsStrain1}
\end{figure}

Simulations of uniaxial compression tests have been performed for values of (single-scale) macro-porosity equal to $\{$5\%, 10\%, 15\%, 20\%, 25\%, 30\%, 35\%, 40\%$\}$ (see the supporting material).
For each value of porosity, ten specimens with different distributions of pores have been considered and,  
accordingly, ten different overall stress/strain responses have been obtained [in terms of $\bar{\sigma} = \mbox{(measured force)}/\textup{(cross-section area)}$ and $\bar{\varepsilon} = \mbox{(applied displacement)}/$L]. Results for three (of the ten) samples characterized by a macro porosity of 20\% are shown in Fig. \ref{StressvsStrain1}.

Here, in full agreement with experimental results, simulations show that the response of the specimens remains almost linear until a peak value of stress is reached, and it is followed by a softening branch, due to the coalescence of small cracks. 

The shape of the softening branch and the value of the peak stress greatly depend on the geometrical distribution of voids, while the stiffness (the tangent line to the stress-strain curve) is only slightly influenced. 
Fig. \ref{StressvsStrain1} is enhanced 
with insets, showing crack growth  at different test stages, developing parallel to the direction of applied loading. The last feature, agreeing with experiments, reveals that the model can effectively capture damage nucleation, fracture propagation and the final failure mechanism.

The mean value and standard deviation of overall stiffness (Young's modulus $\bar{E}$) and  strength (peak stress 
 $\bar{\sigma}$) are reported in  Table \ref{table:3}, evaluated using simulations pertaining to the ten specimens tested for each porosity level. 
Predictions obtained with Eqs.\eqref{eq:coble} and \eqref{eq:strength} 
are included in parentheses and show a
tight agreement with the  numerical values.
\begin{table}[h!]
\centering
\scalebox{0.85}{
\begin{tabular}{ c c c c c c c c c } \hline
  & $\Phi_{\textup{macro}}=5\%$ & $10\%$ & $15\%$ & $20\%$ & $25\%$ & $30\%$ & $35\%$ & $40\%$ \\ \hline
Mean($\bar{E}$)& 61.9 (63.2) & 55.2 (56.7) & 49.1 (50.6) & 43.5 (44.8) & 38.9 (39.4) & 34.4 (34.3) & 29.8 (29.6) & 25.2 (25.2) \\
Mean($\bar{\sigma}$)& 4.53 (5.38) & 3.34 (3.99) & 2.76 (2.96) & 2.13 (2.19) & 1.76 (1.62) & 1.53 (1.20) & 1.25 (0.89) & 1.04 (0.66)\\
St.\,Dev($\bar{E}$) & 0.17 & 0.22 & 0.79 & 1.06 & 1.15 & 1.29 & 0.46 & 1.57\\
St.\,Dev($\bar{\sigma}$)  & 0.17 & 0.17 & 0.23 & 0.18 & 0.21 & 0.18 & 0.08 & 0.12
\end{tabular} }
\caption{\textbf{
Overall properties of compressed 
Biomorphic Apatite samples (generated in silico), 
 where only one scale of porosity is considered.} Mean value (denoted as \lq Mean') and standard deviation (denoted as \lq St.\,Dev') of overall 
Young's Modulus $\bar{E}$ and peak stress $\bar{\sigma}$ for different values of macro-porosity, as obtained from simulations of compression tests. Quantities are expressed in MPa. Values obtained with the  Coble-Kingery, eq. (\ref{eq:coble}), and Rice, eq. (\ref{eq:strength}), are reported in parentheses. 
}
\label{table:3}
\end{table}
It can be noted from the table that at an increase of macro-porosity, the effect of the void distribution on the overall strength tends to decrease, while the stiffness remains almost constant.

\begin{figure}[ht]
\centering
\includegraphics[width=0.65\linewidth]{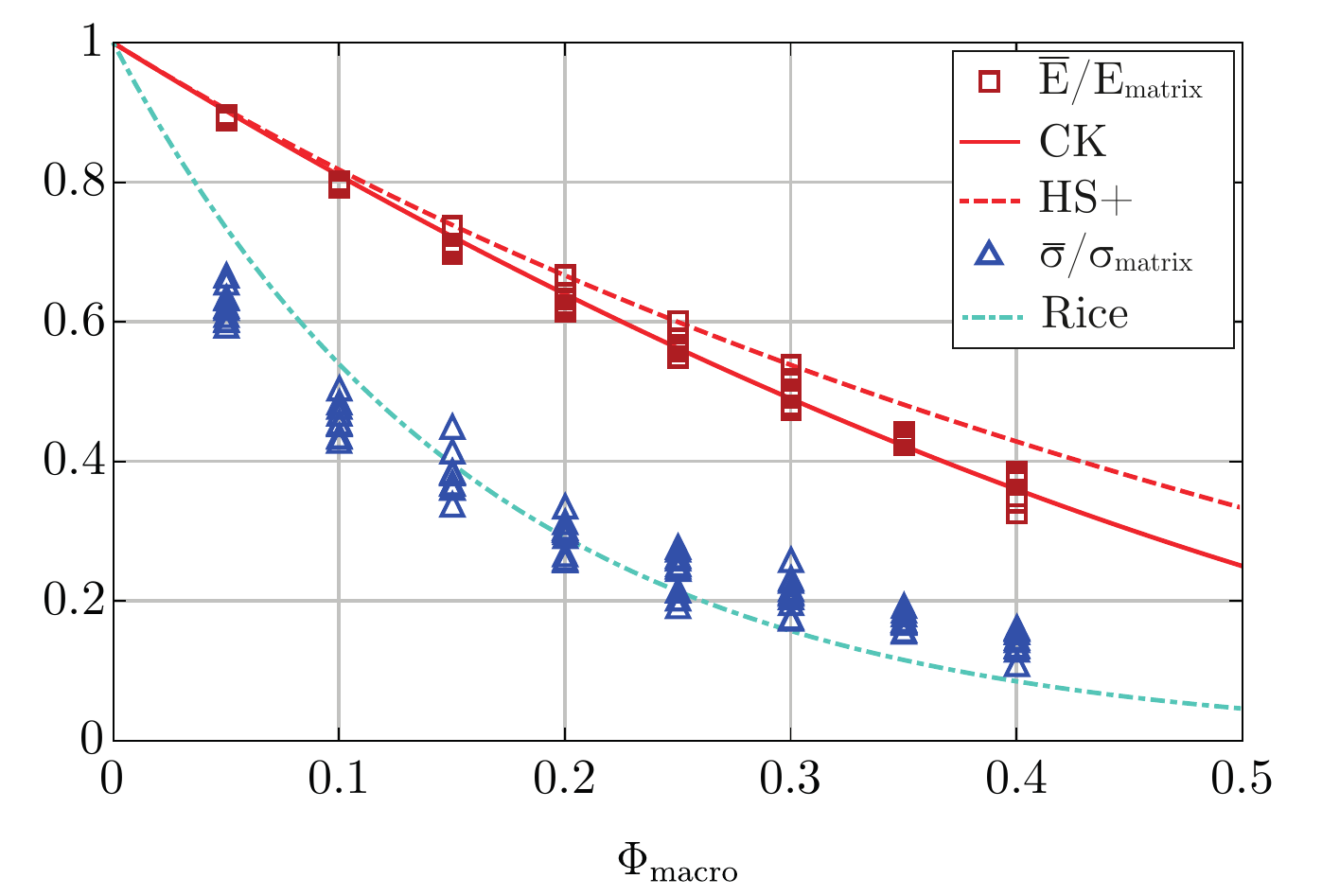}
\caption{\textbf{Overall response of in-silico generated samples of 
Biomorphic Apatite, where only one scale of porosity is considered.} 
Simulated overall peak stress $\bar{\sigma}$ (triangular spots) and overall Young's Modulus $\bar{\textup{E}}$ (square spots) 
as functions of macro-porosity, $\Phi_{\mbox{macro}}$. The lines, reporting the Hashin-Shtrikman upper bound (HS$+$), the Coble-Kingery (CK) and Rice formulas show an excellent agreement.}
\label{EvsSigma2}
\end{figure}

Fig. \ref{EvsSigma2} summarizes results for  specimens 
characterized by a single-scale porosity, 
by displaying values of peak stress and stiffness, as functions of the amount of porosity. Data compare well with predictions from Eq.\eqref{eq:coble}, from the Hashin-Shtrikman upper bound, and Eq.\eqref{eq:strength}. 

\subsection{Compression test on porous brittle solids: two-scale porosity}
\label{doubleimpact}

The phase-field approach to fracture mechanics is shown now to answer several questions that may arise when modeling or designing functional porous ceramics. 
In particular, at a fixed value of total porosity
(7 values are explored, see supplementary material), 
simulations on in-silico created samples 
are used with different percentages of macro and meso voids, to  
explore effects related to second-scale porosity on:  
(i.) overall Young's modulus, (ii.) failure stress, (iii.) crack growth, and (iv.) damage diffusion. 

\begin{figure}[ht]
\centering
	\includegraphics[width=\textwidth]{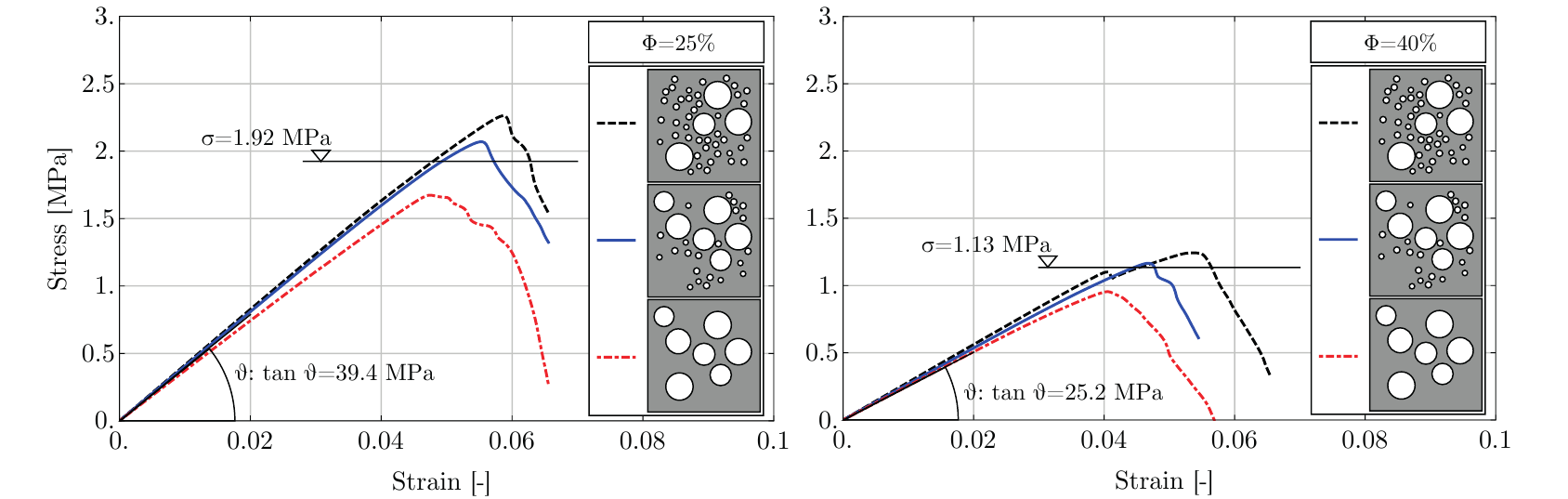}
\vspace{-5mm}
\caption{\textbf{Influence of multiscale porosity on the overall behaviour of Biomorphic Apatite samples (generated in silico).} The effects related to the presence of a meso-pores is visible in the 
simulated overall stress-strain responses for samples characterized by the same value of total porosity $\Phi$ ($25\%$ on the left and $40\%$ on the right), but different amounts of meso $\Phi_{\mbox{meso}}$ and macro $\Phi_{\mbox{macro}}$ porosity. 
The Young's modulus $\bar{E}=\tan\theta $ and peak stress $\bar{\sigma}_{\mbox{peak}}$, Eqs.\eqref{eq:coble} and \eqref{eq:strength}, are included.
Peak stress and elastic stiffness increase when the density of meso-pores is increased.
 }
\label{stressstrain2scale}
\end{figure}

Overall stress/strain curves, analogous to those presented in Fig. \ref{StressvsStrain1}, but now computed for double-scale porosity, are depicted in Fig. \ref{stressstrain2scale}. 
Here, the total porosity (sum of the macro and meso values, $\Phi=\Phi_{\textup{macro}}+\Phi_{\textup{meso}}$) is assumed to be equal to 25\% (left) and 40\% (right). Each curve represents the response of a sample in which the overall porosity is achieved by mixing different portions of macro-pores and meso-pores, where the former pores have a larger size than the latter, according to the PDF, Eq.\eqref{eq:PDF}.
The peaks of the curves shown in 
Fig. \ref{stressstrain2scale} reveal that a superior mechanical response for specimens 
follows from an increase in the percentage of meso-pores. The left image of the same figure shows specimen with a total porosity of $25\%$ split in macro-porosity and meso-porosity as follow: the red dot-dashed curve is characterised by $25\%$ of macro-porosity, the blue one by $20\%$ macro-porosity and $5\%$ meso-porosity, and lastly, the black dashed one by $15\%$ macro-porosity and $10\%$ meso-porosity. The image on the right shows the effect is reproduced also for a total porosity of $40\%$.

\begin{figure}[ht]
\centering
\includegraphics[width=0.6\linewidth]{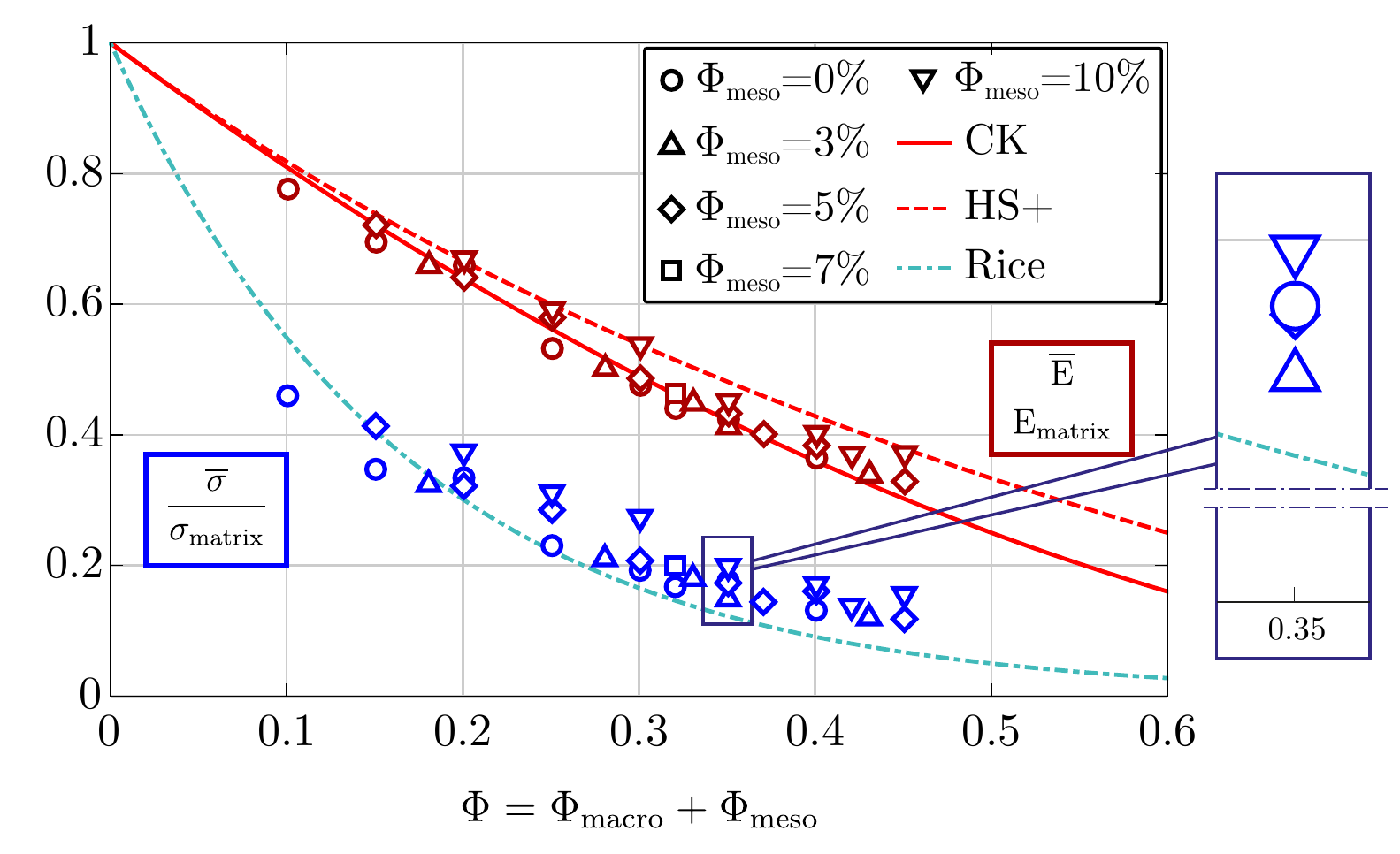}
\vspace{-2mm}
\caption{\textbf{Overall properties of 
Biomorphic Apatite samples (generated in silico), 
with pores of different sizes.} Simulated dimensionless Young's modulus (red) and 
peak stress (blue) versus total porosity for samples characterized by double porosity. The inset shows that at $\Phi=0.35$, the presence of a 0.03 fraction of meso-porosity can (surprisingly and contrary to most of the results) lead to a smaller stiffness and strength than the case where meso-porosity is absent. This is related to the specific geometrical distribution of voids.
}
\label{specifiche_provini_2scale2}
\end{figure}

Overall properties, in terms of peak strength and elastic modulus, for different specimens are shown in Fig. \ref{specifiche_provini_2scale2}, as functions of the total porosity, but for different 
values of meso (and thus also macro) porosity. 
According to \cite{Cui2020}, the figure reveals that for a fixed value of total porosity, samples display improved mechanical properties at low macro porosity (triangles are located clearly above the trend lines). Therefore, macro porosity is more detrimental to stiffness and strength than meso-porosity. However, our computations show that this is not a strict rule, as evidenced in the inset of Fig. \ref{specifiche_provini_2scale2}, showing that a sample at 0.35 total-porosity with 0.03 meso-porosity is less stiff and resistant than the sample with null meso-porosity. This feature is not isolated and occurs also at 
$\Phi = 0.2$ and is 
related to the particular geometrical distribution of voids and prompts the idea that the design of the spatial void distribution (possible via phase field) could improve material performances.

\begin{figure}[ht]
\centering
\includegraphics[width=\linewidth]{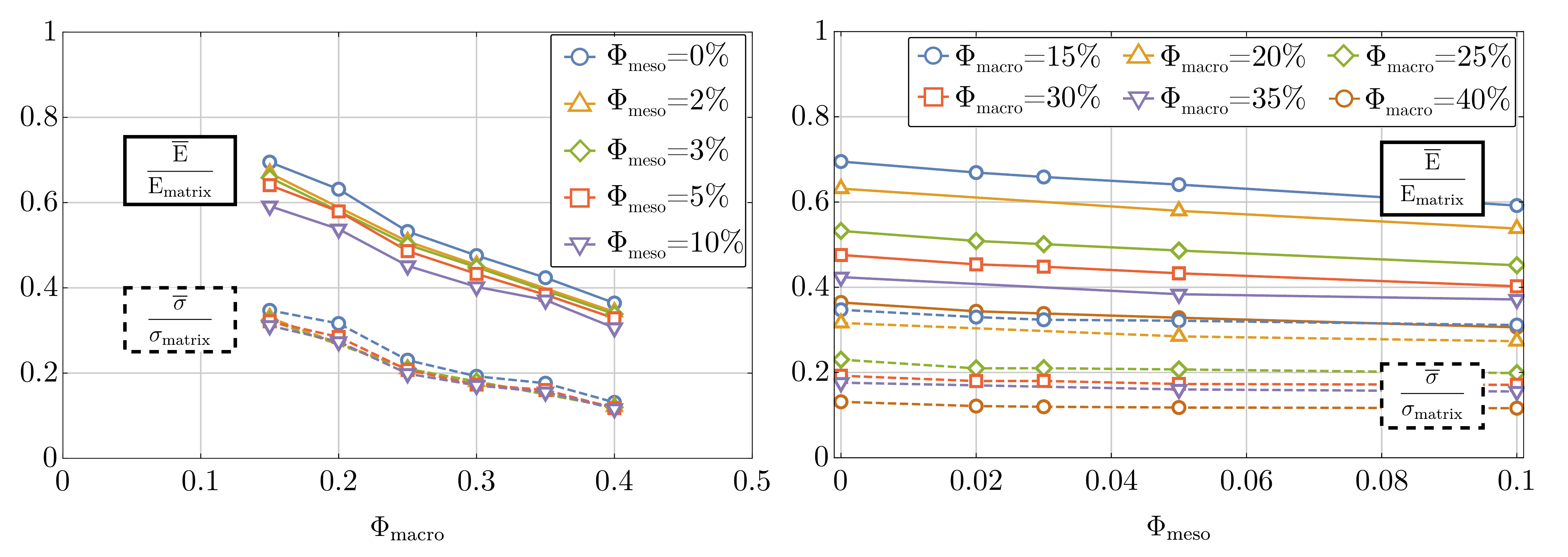}
\vspace{-5mm}
\caption{\textbf{Dependence of the mechanical behaviour of 
Biomorphic Apatite 
 on the voids' size distribution.} Effects of the meso porosity fraction on the  dimensionless peak stress and Young's modulus as a function of the macro porosity (left) and vice-versa ( right). 
}
\label{fig:2scale_variability}
\end{figure}
\begin{figure}[ht]
\centering
\includegraphics[width=0.85\linewidth]{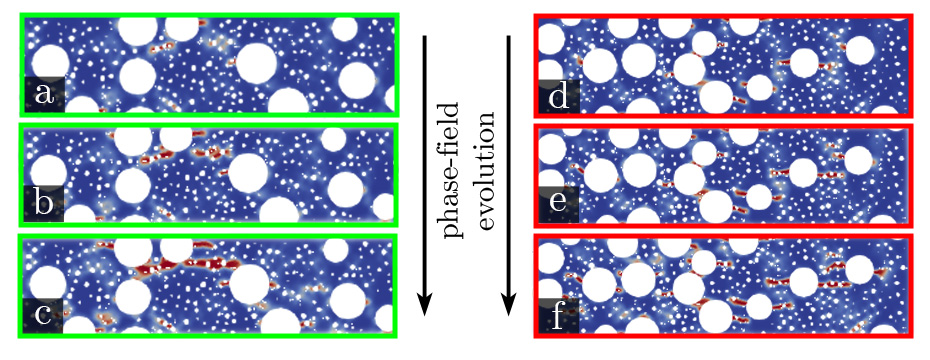}
\caption{\textbf{Crack evolution in 
Biomorphic Apatite 
with multiple scale porosity using the phase-field.} Simulated fracture evolution through phase-field during compression of a sample characterized by a macro-porosity $\Phi_{\textup{macro}}=25\%$ and meso-porosity $\Phi_{\textup{meso}}=10\%$. Meso-pores induce two failure  mechanisms, either involving nucleation (a-b) and growth (c) of cracks, or connecting macro-pores (d-e) and ultimately leading to failure (e).}
\label{samplesAPF}
\end{figure}

Both scales of porosity are found to concur in producing a non-linear effect on the overall properties of the porous material. In particular, an interaction is observed between the two scales of porosity (the sum between the two $\Phi_{\rm macro}$ and $\Phi_{\rm meso}$ is constant and equal to $\Phi$). This emerges from the plots of the dimensionless 
peak strength and Young's modulus 
reported in Fig. \ref{fig:2scale_variability}
as resulting from simulated stress-strain curves, functions  of macro and meso-porosity $\Phi_{\rm macro}$ and $\Phi_{\rm meso}$. 
Simple linear regression of the normalized overall properties obtained from our two-scale porosity analyses (depicted in Fig. \ref{fig:2scale_variability}) leads to the following approximations
\begin{eqnarray}
    \frac{\overline{\textup{E}}}{\textup{E}_{\textup{matrix}}}&=&0.905-1.401 \Phi_{\textup{macro}}-0.868\Phi_{\textup{meso}}, \ \textup{R}^2=0.913 ,\\
    \frac{\overline{\sigma}}{\sigma_{\textup{matrix}}}&=&0.610-1.303\Phi_{\textup{macro}}-0.704\Phi_{\textup{meso}},\ \textup{R}^2=0.748 ,
\end{eqnarray}
where R$^2$ is the coefficient of determination.\\
Crack patterns are found to be strongly influenced by meso-porosity at fixed values of macro-porosity, as shown in Fig. \ref{samplesAPF}, 
where simulations of crack patterns in two portions of a sample (one framed green and the other red) are 
shown at different stages of loading, during a compression test for  $\Phi_{\textup{macro}}=25\%$
and 
$\Phi_{\textup{meso}}=10\%$. 
Here, simulations reveal the existence of 
two mechanisms of failure affecting specimens 
characterized by double porosity: (i.) 
in the portion of the sample framed red,  
cracks nucleate at the edge of macro-pores, while meso-pores promote the development of a crack network, while (ii.) in the portion framed green, cracks nucleate near meso-pores and develop by joining macro-pores.
\\

\begin{figure}[ht]
\centering
\includegraphics[width=1\linewidth]{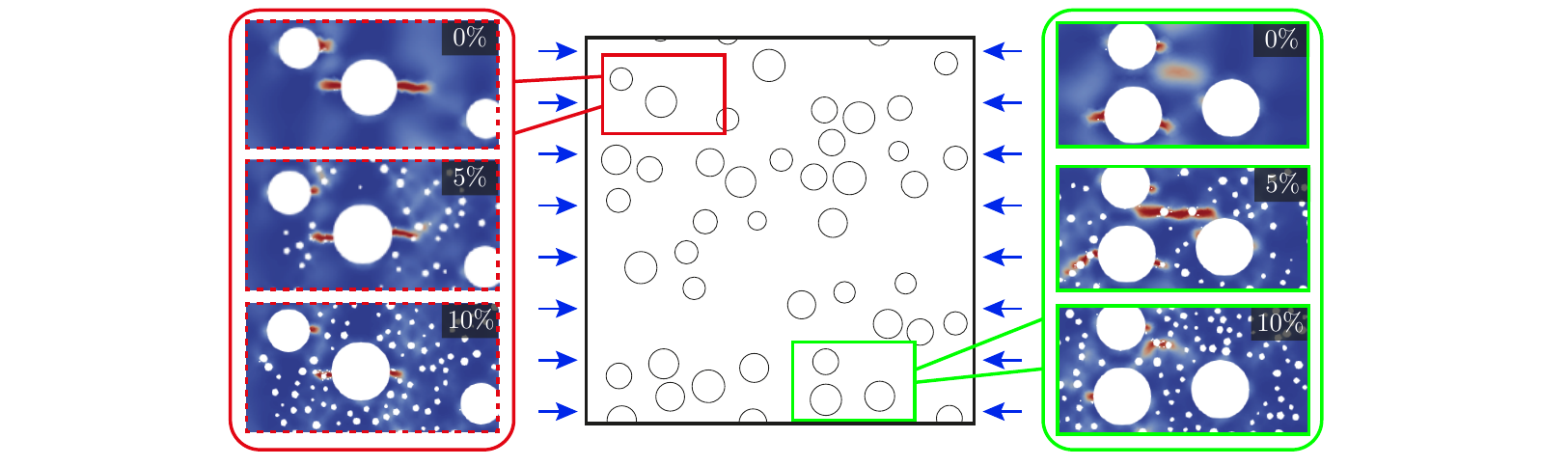} 
\vspace{-5mm}
\caption{\textbf{Effects of voids of small size on crack nucleation and growth in 
Biomorphic Apatite.} 
Cracks develop parallel to the loading direction in the 
simulated fracture evolution through phase field 
during compression of a sample characterized by a macro porosity 20\% and different values of meso-porosity equal to $\{$0\% 5\%, 10\%$\}$. The reported phase-field refers to an overall strain of $\bar{\varepsilon}=-6.1\times 10^{-2}$.
}
\label{fig:samples_B}
\end{figure}

\paragraph{Crack paths in two-scale porous specimens} The presence of meso-porosity does not only affect the overall mechanical behaviour of a specimen in terms of stiffness and strength, but also its failure. 
The meso-porosity enhances damage diffusion and promotes new fracture paths, different from those observed in the absence of meso pores, as visible 
in Fig. \ref{fig:samples_B}, 
on the left, for $\Phi_{\mbox{macro}}=20\%$. Here crack patterns are analyzed 
in a region of a sample compressed 
at a fixed value $\bar{\varepsilon}=-6.1\times 10^{-2}$ of overall strain, but with different amounts  of meso-porosity, equal to $\{$0\,\%, 5\,\%, 10\,\%$\}$. 
In the region reported on the right, meso-porosity, which produces large volumes of solid where damage grows, is shown to shield from damage. 

\begin{figure}[ht]
\centering
\includegraphics[width=0.7\linewidth]{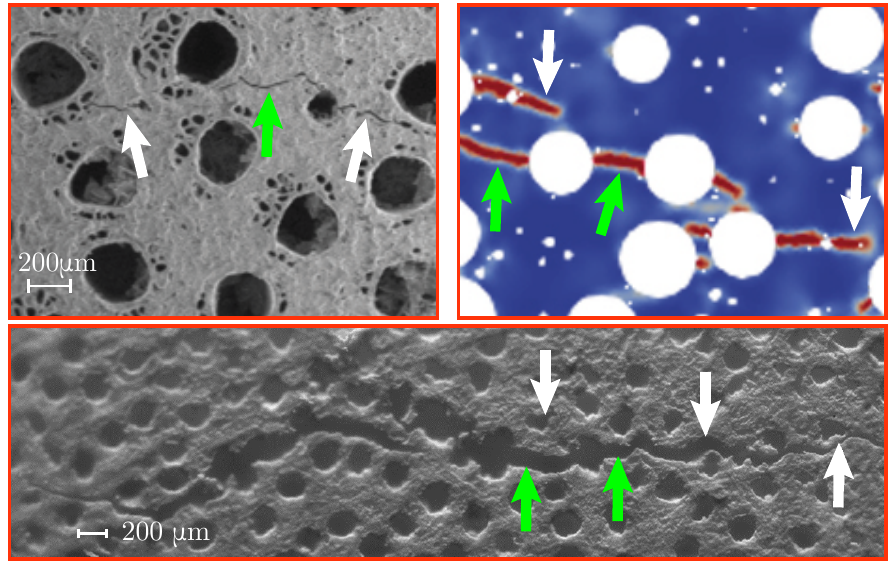} 
\vspace{-1mm}
\caption{\textbf{Phase-field successfully reproduces crack propagation mechanisms in 
biomorphic apatite.} Upper part on the left and lower part: experiments show 
crack propagation nearly parallel to the applied compressive stress. The same feature is found in the simulation on a randomly generated sample (thus not identical neither to the sample on the left nor to that in the lower part). The simulation was terminated at an overall strain of $6.5\times 10^{-2}$. The photo in the upper part was made with a SEM, while the other  
 with a confocal microscope (Leica DCM3D), during a compression test. Note that the tortuosity of the crack paths is the result of almost straight cracks connecting macro-pores (green arrows), while inclined cracks originate from micro-pores (white arrows).}
\label{crackBA}
\end{figure}

A direct comparison between a phase-field simulation of fracture growth in a porous material and two micrographs obtained during a compression test on BA, performed at the MUSAM-Lab (financed with 
the ERC StG CA2PVM, Grant Agreement 306622, and the ERC PoC PHYSIC, Grant Agreement No. 737447, at IMT School, Lucca), is presented in Fig. \ref{crackBA} (upper part on the left and lower part). 
The tests were executed by placing prismatic samples in direct contact with two steel platens (lateral size $40\ \mathrm{mm}$ and thickness $5\ \mathrm{mm}$) and compressed (using the tensile/compressive stage DEBEN 5000S inside the scanning electron microscope Zeiss EVO MA15) by imposing displacements. The test was continued until complete failure of the samples. Two mechanisms of crack propagation can be observed in Fig. \ref{crackBA}: cracks either nucleate in the proximity of large pores and develop horizontally (highlighted in the figure with green arrows), or they nucleate near meso pores, thus producing a fracture network, where macro and meso pores are connected
(highlighted in the figure with white arrows). 
The results of the simulation (upper part on the right)
are not {\it coincident} with 
the fracture experiments because the sample used for simulation is randomly generated and therefore not identical to those analyzed experimentally, but the mechanisms of crack nucleation and growth are the same. 

Note that, while the test reported in the upper part of Fig. 
was performed inside the SEM, that reported in the lower part of the figure was conducted in air with the same tensile/compressive stage, but examined with a confocal microscope (Leica DCM3D) with a magnification lens 10$\times$. In this way, a larger portion of the sample was observed, and one typical result is shown in Fig.\ref{crackBA} (lower part). Again, sub-horizontal cracks are observed (green arrows), connecting macro pores, while inclined cracks nucleated near meso pores and later coalesced to join a \lq main' propagating crack, evidencing a characteristic tortuosity. Cracks originated from meso-pores can sometimes shield macro-pores, as clearly indicated by the white arrows in Fig.\ref{crackBA}.
All these features are accurately reproduced in the simulations. 

\pagebreak

\section{Discussion}

Simulations carried out with the phase-field approach to fracture on samples with cylindrical porosity (generated in-silico to model a porous ceramics obtained from wood, biomorphic apatite) show that the presence of two size-scales of porosity plays an important role in damage diffusion and fracture growth.
In particular, the simulations (performed with an in-house developed finite element code, 
calibrated on brittle materials, available in the Supplementary materials)  show that during compressive failure:
(i.) fractures nucleate at pore boundaries and grow parallel to the direction of compression; (ii.) 
large pores induce crack nucleation, while small pores foster their propagation by connecting distantly located voids; (iii.) small pores cause the formation of micro-cracks, eventually promoting and propagating failure; (iv.) small pores diffuse damage and reduce the inclination of the softening branches in the overall stress-strain curves. 

At a given value of the total porosity, specimens characterized by high values of meso-porosity display mechanical properties superior than those characterized by a small number of large pores. Therefore, the pore size distribution has to be considered a major influential parameter in the mechanical design of porous brittle materials, where the phase-field approach reveals its excellent potentialities.

\pagebreak

\section*{Data availability}
\noindent All the raw/processed data required to reproduce these findings are provided in the Appendix A, both in terms of finite element codes to generate and run the computational models, and in terms of key research results.

\section*{Acknowledgements}
D.B. and R.C. acknowledge financial
support from the European Research Council Advanced Grant scheme (ERC-AdG), under the European Union’s Horizon 2020 research and innovation programme (Grant agreement No. ERC-ADG-2021-101052956-BEYOND). 

Experiments were performed (i.) at 
MUSAM (IMT Lucca) using equipment purchased with the support of the grants: ERC StG CA2PVM (Grant Agreement 306622), ERC PoC PHYSIC (Grant Agreement No. 737447), (ii.) at 
the Instabilities Lab (University of Trento), 
and (iii.) at  the 
Laboratory for Numerical Modelling of 
Materials (University of Trento), using equipment purchased with the support of the grants: ERC-ADG-2021-101052956-BEYOND and
ERC-2013-ADG-340561-INSTABILITIES. 

This work has been partially supported by the Italian Ministry of University and Research (MUR) through the
project `Scientific computing for natural sciences, social sciences, and applications: methodological and technological development' (PRO3 Joint Program 2022-23, CUP D67G22000130001), which is gratefully acknowledged by P.L. and M.P.\\
R.C. acknowledges financial support from project \lq Studying the Colloidal Effect on the Spreading of Carbon-Based Agents on Hair' funded by \lq Procter \& Gamble Company'.

\bibliographystyle{elsarticle-num}
\bibliography{reference}

\end{document}